\documentclass[conference]{IEEEtran}
\IEEEoverridecommandlockouts
\usepackage{cite}
\usepackage{amsmath,amssymb,amsfonts}
\usepackage{algorithmic}
\usepackage{graphicx}
\usepackage{textcomp}
\usepackage{xcolor}
\usepackage{balance}
\usepackage{hyperref}
\def\BibTeX{{\rm B\kern-.05em{\sc i\kern-.025em b}\kern-.08em
    T\kern-.1667em\lower.7ex\hbox{E}\kern-.125emX}}
\begin{document}

\title{The Dependency Black Hole

\thanks{This research was supported by KK foundation through the KK-H\"{o}g project 2024/0140 and KKS Profile project SERT 2018/010 at Blekinge Institute of Technology, Sweden.}
}

\author{\IEEEauthorblockN{Ehsan Zabardast\IEEEauthorrefmark{1}\IEEEauthorrefmark{2}, Bhuwan Paudel\IEEEauthorrefmark{1}, Javier Gonzalez-Huerta\IEEEauthorrefmark{1}}
\IEEEauthorblockA{
\textit{\IEEEauthorrefmark{1}Software Engineering Research Lab (SERL)}\\
\textit{Blekinge Institute of Technology},
Karlskrona, Sweden\\ 
\{ehsan.zabardast, bhuwan.paudel, javier.gonzalez.huerta\}@bth.se\\
}
\IEEEauthorblockA{\IEEEauthorrefmark{2}\textit{Gaetir}, 
Karlskrona, Sweden}}

\maketitle

\begin{abstract}

Microservice architectures promise independent evolution through loose coupling, yet large systems often exhibit strong dependency concentration around a small set of services. In an exploratory industrial case study of a product composed of 267 microservices, we triangulated multiple dependency signals---compile-time, run-time, and task dependencies---and iteratively validated our interpretations with practitioners. We observed a recurring macro-structure in the dependency network that resembles a black hole: a dense core of dependency magnets, a transitional region of services increasingly entangled with the core, and an outer region of lightly connected services. Based on these observations, we propose the \emph{dependency black hole} theory, mapping the network to the black hole anatomy of a singularity, an event horizon, and an accretion disk, and formulating three hypotheses about how dependency concentration emerges and evolves at scale. The theory provides an explanatory lens for reasoning about dependency growth, identifying services at risk of becoming dependency magnets, and motivating governance interventions. We outline practical implications and directions for longitudinal and multi-case validation.

\end{abstract}


\section{Introduction}\label{sec:introduction}

The development of software-intensive products has increasingly moved towards ``componentising'' architectures in which functionality is decomposed into smaller, independently deployable units. A prominent example is the microservices architectural style, which promises agility and evolvability through decentralisation and the isolation of concerns. Yet, microservices are rarely truly independent: they interact to deliver end-to-end value, and these interactions form \emph{architectural dependencies}. In this paper, we use the term architectural dependency to denote any relationship in which a microservice becomes constrained by another microservice for correct behaviour and evolution. Consequently, dependencies strongly influence change impact, fault propagation, governance needs, and architectural sustainability.

While dependency management is non-trivial in any system, the challenges become acute at scale in microservice-based architectures~\cite{zabardast2025architecture}. As the number of services grows, the dependency network becomes harder to observe and reason about, especially when dependencies are scattered across heterogeneous sources (code, deployment descriptors, run-time traces, and organizational artifacts). Empirical evidence suggests that architectural complexity indicators---including dependencies and inter-service communication---tend to co-evolve and increase as microservice systems evolve~\cite{paudel2025temporal}. In parallel, studies of microservice anti-patterns and bad smells repeatedly report problematic dependency structures (e.g., excessive coupling, improper boundaries, and hidden shared resources) in practice~\cite{cerny2023catalog}. These issues are not purely technical: they can amplify coordination costs and contribute to architectural technical debt~\cite{de2021identifying,BOROWA2025112547}.

This trajectory is consistent with Lehman's laws of software evolution: systems that remain useful must undergo \emph{continuous change}, and their complexity tends to \emph{increase} unless explicit work is performed to control it~\cite{lehman1979understanding}. In large microservice systems, dependencies are a primary mechanism through which such growth in complexity becomes visible and consequential. Despite the growing body of knowledge on microservice smells and complexity evolution, there is still a limited theory-driven understanding of \emph{how} dependency concentration emerges in large microservice dependency networks and why a small subset of services tends to dominate the dependency landscape over time.

In this paper, we present the \emph{dependency black hole} as a theory-building contribution aimed at explaining a recurring macro-structure we observed in a large-scale industrial microservice system. The theory characterises dependency networks using an analogy to the anatomy of a black hole---\emph{singularity}, \emph{event horizon}, and \emph{accretion disk}---to describe distinct ``regions'' of the architecture in which services have different dependency dynamics and risks of further entanglement. We grounded the theory in an industrial case study with $267$ microservices, synthesising multiple dependency data sources including compile-time, run-time, and task dependencies, and iteratively validating our interpretation with engineering managers, product managers, and developers.

The remainder of the paper is organised as follows. Section~\ref{sec:industrial_case_study} describes the industrial case study design, data sources, and processing pipeline. Section~\ref{sec:dependency_black_hole} introduces the dependency black hole theory and its core hypotheses. Section~\ref{sec:validation} reports the initial validation with practitioners. Section~\ref{sec:discussion} discusses implications for architectural governance and practical dependency management. Section~\ref{sec:related_work} positions the work within existing research on microservice dependencies, coupling, and dependency magnets. Finally, Section~\ref{sec:conclusion} concludes and outlines future work on longitudinal validation across further cases.

\section{Industrial Case Study}\label{sec:industrial_case_study}

We have conducted an exploratory industrial case study in close collaboration with an industrial partner. The study follows established case study guidelines for software engineering and uses methodological triangulation by combining quantitative dependency data with qualitative feedback from practitioners~\cite{ralph2020empirical,runeson2012case}.

\textbf{1) Goal:} The goal is to study architectural dependency complexity in a large-scale microservice architecture and to develop an initial explanatory theory about how dependencies concentrate and change as the system evolves. In particular, we aim to (i) characterise dependency structures across microservices, (ii) identify recurrent dependency patterns (e.g., potential dependency magnets, meaning services that accumulate disproportionate numbers of direct and transitive dependencies, becoming an over-centralised services in the architecture), and (iii) iteratively refine a theory grounded in observations and practitioner discussion.

\textbf{2) Sample and Population:} The case system is a large industrial product developed using a microservice architecture. The company has chosen to remain anonymous. Our population consists of $267$ microservices that collectively deliver the product's end-to-end functionality. The microservices are developed and maintained by multiple autonomous teams operating with agile and DevOps practices.

\textbf{3) Design:} We employed an iterative analysis-and-feedback design. We repeatedly (a) extracted and synthesised dependencies and visualisation, (b) analysed the resulting network structures, and (c) discussed interim interpretations---every two weeks---with the teams. These iterations were used to challenge, refine, and consolidate the emerging theory, and to ensure that the identified dependency structures were meaningful in the practitioners' context.

\textbf{4) Data Collection and Processing:} We collected dependency-related data from multiple sources to build a multi-perspective dependency network. The data includes: (i) compile-time and run-time dependencies (e.g., libraries and transitive dependencies inferred from build and repository metadata); (ii) task dependencies derived from issue-tracking data (e.g., Jira tickets linking work items to microservices and teams); (iii) team affiliation and contribution signals; and (iv) microservice ownership data. Data was collected via the partner's tool APIs, then pre-processed to remove incomplete or inconsistent records (e.g., missing service identifiers, ambiguous ownership entries, and duplicate links). Finally, the cleaned datasets were used to generate dependency graphs that were used as input to the analysis.

\section{The Dependency Black Hole}\label{sec:dependency_black_hole}

\begin{figure*}[htbp]
\centerline{\includegraphics[width=1\textwidth]{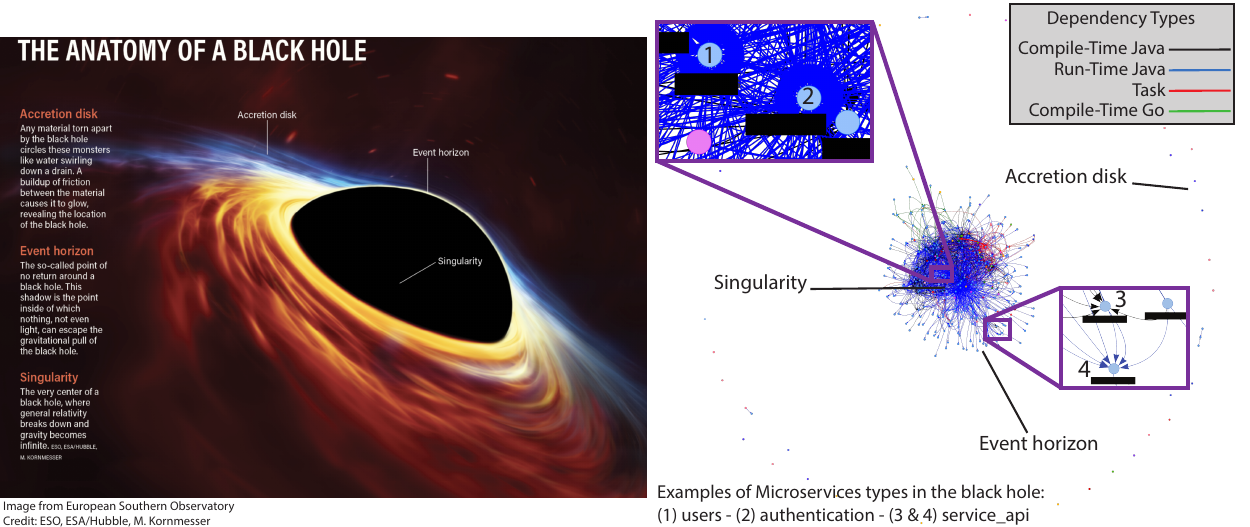}}
\caption{Visualization of the dependency network and its resemblance to the black hole anatomy. Dots represent a microservice (colours signify owners), and lines represent dependency between microservices (colours signify the type of dependency).}
\label{fig:dependency_black_hole}
\end{figure*}

Based on our observations of the case, the data we collected, and our continuous discussions with managers and developers (Section~\ref{sec:validation}), we developed the dependency black hole theory. When visualising dependencies between microservices as a directed network, we observed a characteristic macro-structure: a small set of services accumulates an outsized share of incoming and transitive dependencies, while most services remain comparatively peripheral. The resulting shape resembles a black hole (see Fig.~\ref{fig:dependency_black_hole}): a dense core surrounded by a transition zone and an outer region of lightly connected services. We use the anatomy of a black hole as an explanatory analogy that provides vocabulary for reasoning about dependency concentration and its evolution.

\subsection{Regions of the Dependency Black Hole}
\textbf{Singularity} consists of \emph{dependency magnets}: services that become highly central in the architecture because many other services depend on them directly or indirectly, causing them to repeatedly appear on dependency paths across the system. These services typically offer cross-cutting capabilities or shared resources, such as identity and access rights, core platform services, shared integration endpoints, or canonical data/contract services. In our case, practitioners repeatedly pointed to such services as ``unavoidable'' integration points, often justified by standardisation and reuse. However, their centrality increases the blast radius of change and failure, and can lead to architectural bottlenecks and coordination overload.

\textbf{Event horizon} is a transitional region populated by services that are not dependency magnets themselves, but have established non-trivial dependencies \emph{towards} the singularity (directly or through short transitive chains). These services tend to mediate flows to or from the magnets (e.g., aggregation, orchestration, adaptation, or domain gateways). Our key observation is that services in this region are at risk of crossing a threshold and becoming magnets over time as additional responsibilities, integrations, or ownership changes accumulate and the number of dependencies tends to grow over time~\cite{paudel2025temporal}.

\textbf{Accretion disk} comprises services with few dependencies, often implementing more isolated domain functionality, bounded context-specific logic, or niche capabilities. They may depend on other peripheral services, but have no (or minimal) reliance on the singularity. In practice, these services are frequently easier to change and deploy independently, and they contribute less to system-wide coordination pressure.

\subsection{Hypotheses and Illustrative Examples (Initial Results)}
The dependency black hole theory is captured by three hypotheses derived from the case observations. 

\textbf{H1:} \textit{Large-scale microservice-based systems tend to form a dependency black hole once they surpass a threshold.}

As a microservice system grows (in number of services and integrations), dependency relations increasingly concentrate around a small subset of services, forming a dense core (singularity) that dominates the dependency network.

In our case, the dependency distribution was markedly skewed: most services had only a small number of dependencies, while a small subset accumulated substantially more incoming and transitive dependencies than the median service. The practitioners identified these highly central services as cross-cutting capabilities that the majority of services ``must'' interact with early in their lifecycle. Examples include identity and access services (e.g., authentication/authorisation), shared platform services (e.g., service registry/configuration), and canonical integration endpoints. This aligns with a cumulative advantage mechanism: once a service becomes a common integration point, it becomes increasingly likely that future services will depend on it as well.

\textbf{H2:} \textit{Services in the event horizon are vulnerable to ``falling'' into the singularity as dependencies accumulate.}

Services that establish several dependencies towards magnets (and become intermediaries for multiple flows) may cross a dependency threshold and transition into dependency magnets themselves. We observed multiple services that initially acted as thin adapters or orchestrators around central services (e.g., mediating between domain functionality and identity or a platform capability), but gradually expanded into broader coordination roles. In the dependency graphs, these services were characterized by (i) short-path proximity to the magnets and (ii) repeated appearance on dependency paths between peripheral services and the singularity.

\textbf{H3:} \textit{Many services remain in the accretion disk with few dependencies and limited exposure to the singularity.}

A substantial fraction of services remains lightly connected and relatively isolated, typically depending on only a few services and often not directly connected to the singularity.

In our case study, many services formed a sparse outer region with few direct dependencies and limited transitive reliance on the magnets. Practitioners characterised these services as bounded-context implementations that could evolve relatively independently. In the dependency graphs, these services tended to have low fan-in and low centrality, and were more likely to interact via asynchronous messaging or through domain-local dependencies rather than directly calling core magnets. Notably, several accretion-disk services exhibited small clusters (local neighbourhoods) with dependencies among themselves, suggesting that micro-level modularity can still exist even when macro-level concentration emerges elsewhere.

\subsection{Why the Analogy Matters?}
The black-hole analogy is not merely visual: it provides a compact system-level explanation of how dependency concentration can emerge and why it is difficult to reverse. Once a service enters the singularity, architectural change becomes more expensive due to widespread dependents, and the system becomes prone to further entanglement as new services choose the ``path of least resistance'' by integrating with existing magnets. Similarly, the event horizon highlights a critical governance zone where architectural interventions (e.g., stricter contract management, ownership rules, and dependency budgeting) may prevent additional services from transitioning into magnets.

\section{Initial Validation}\label{sec:validation}

Given the exploratory and theory-building nature of this work, we performed an initial validation through iterative practitioner feedback. Following each analysis iteration, we presented the evolving dependency network visualisations and intermediate interpretations to key stakeholders, including engineering managers, product managers, and senior developers with long-term knowledge of the system. The sessions focused on (i) assessing whether the extracted dependency structures were plausible and complete (e.g., no major services or relations missing) and (ii) confirming the interpretation of high-dependency services as dependency magnets.

To reduce the risk of confirmation bias, we structured the discussions around concrete artifacts (graphs, ranked service lists, and example dependency paths) and asked participants to identify counterexamples and alternative explanations (e.g., whether observed concentration could be an artifact of the extraction pipeline or naming/ownership inconsistencies). Disagreements were used as triggers for revisiting the data processing steps (e.g., service identifier mapping, deduplication rules, and dependency-type inclusion criteria). Across sessions, practitioners consistently recognised a small set of services as central coordination points (e.g., cross-cutting platform or identity-related services) and acknowledged that new services tend to integrate with these hubs early, supporting the plausibility of the proposed theory.

The outcome of this initial validation is not a statistical proof, but an increased confidence that (i) the constructed dependency network reflects relevant architectural relations in the case system, and (ii) the dependency black hole provides a useful explanatory lens that aligns with practitioners' experience and can guide the formulation of testable hypotheses for future longitudinal and multi-case validation.

\section{Discussion \& Practical Implications}\label{sec:discussion}

Our observations suggest that dependency concentration in large-scale microservice systems is not accidental but a recurring architectural dynamic consistent with continuous change and increasing complexity as described by Lehman~\cite{lehman1979understanding}. Existing evidence on co-evolution of architectural complexity indicators in microservices further supports that dependency-related measures tend to increase as systems grow~\cite{paudel2025temporal, apolinario_method_2021}. In this context, the dependency black hole offers a practical lens for reasoning about where dependency-related risks accumulate and where interventions are likely to be most effective.

\paragraph{Accretion disk: localised issues and low blast radius}
Services in the accretion disk are largely isolated repositories with predominantly technical dependencies (e.g., limited compile-time or small-scale service-to-service links). As a result, architectural problems (e.g., minor coupling or local smells) are usually easier to address within a single team and are less likely to propagate across the system. This aligns with observations that microservice issues are often manageable when coupling remains limited~\cite{cerny2023catalog}.

\paragraph{Event horizon: manageable coupling, but rising risk}
Services in the event horizon have non-trivial dependencies towards central services but are not yet dependency magnets. In our case, these repositories were often owned by the same team or a small set of closely collaborating teams, which reduces coordination overhead and keeps remediation effort relatively low. The blast radius of problems is typically smaller than in the singularity, and fixes can often be implemented with modest effort. However, this region is critical: repeated additions of integration logic and convenience reuse can gradually push services toward magnet-like behaviour, reinforcing the system-wide growth of complexity~\cite{paudel2025temporal,lehman1979understanding}.

\paragraph{Singularity: dependency magnets and high propagation}
Dependency magnets in the singularity have many dependents and appear frequently on transitive dependency paths. Consequently, faults, contract changes, and architectural technical debt can propagate widely, creating large change impact and coordination pressure. This mirrors industrial evidence that architectural technical debt and degradation become especially costly when concentrated around central components~\cite{de2021identifying,BOROWA2025112547}. Practically, singularity services warrant disproportionate architectural attention (e.g., stricter interface governance, stability policies, compatibility strategies, and dedicated ownership).

\paragraph{Socio-technical implications beyond repositories}
A key implication is that the ``entities'' participating in the black hole are not limited to code repositories: ownership and team structures shape how dependencies emerge and how expensive they are to change. This reinforces that architectural dependencies and organisational dependencies co-evolve, and that architecture and organisation go hand in hand when managing microservice complexity at scale~\cite{zabardast2025architecture,tanveer2023approach,de2021identifying}. Moreover, not only the \emph{number} of dependencies but also the \emph{relationship structure} between repositories matters (e.g., transitive paths, concentration, and proximity to magnets), echoing prior calls to consider system-level dependency perspectives when making local changes~\cite{cerny2023catalog,paudel2025temporal}.

\paragraph{Next step - measuring dynamics of services across regions}
An immediate research and engineering next step is to measure the \emph{actual dynamics} of nodes over time: whether, how, and when services approach the event horizon and transition into the singularity. Longitudinal measurement would enable testing the black hole hypotheses as time-dependent claims and would operationalise early-warning indicators (e.g., growth rates of fan-in, centrality, and increasing transitive exposure) for dependency-magnet formation~\cite{paudel2025temporal,lehman1979understanding, apolinario_method_2021, decan2019empirical}.

\paragraph{Additional practical implications}
The theory suggests actionable governance points: (i) treat event-horizon services as a ``control zone'' where dependency budgeting, stricter review of new integrations, and clearer ownership boundaries can prevent new magnets; (ii) invest in observability of dependencies across heterogeneous sources of dependency to avoid blind spots~\cite{zabardast2025architecture}; and (iii) prioritise stabilisation and compatibility engineering for singularity services to reduce propagation and rework costs~\cite{de2021identifying,BOROWA2025112547}.

\section{Related Work}\label{sec:related_work}

Architectural evolution often follows a ``rich get richer'' dynamic, in which services that are already structurally central are more likely to attract additional dependencies over time, leading to the convergence of dependencies on a small set of core services. Recent work has emphasised the importance of dependency structure in understanding microservice architectures. Bakhtin et al.~\cite{bakhtin_network_centrality_ICSA_2025} analyse microservices using a service dependency-graph representation and study the relationship between network centrality and traditional software metrics. Their results show that central microservices tend to expose more public methods and exhibit selective relationships with complexity, inheritance structure, and quality. The study concludes that centrality provides a complementary architectural perspective in identifying the anti-patterns. 

In a subsequent study, Bakhtin et al.~\cite{bakhtin_ccp_ecsa_2025} introduce Centrality Change Proneness (CCP) rank to quantify how likely a microservice is to change its centrality position over time. Their temporal analysis demonstrates that the centrality of microservices can vary substantially across releases, with such temporal instability serving as an early indicator of architectural degradation~\cite{arch_degradation_li_sms}. Likewise, in a recent longitudinal study, Paudel et al.~\cite{paudel2025temporal} investigated the temporal evolution of architectural complexity indicators. The finding revealed that these indicators co-evolve and consistently increase as the system expands. Additionally, the study notes that a small subset of services can skew the overall behaviour of complexity indicators (such as dependency growth). 

Assunção et al.~\cite{ASSUNCAO_msa_evolution} analyse $7,319$ commits across $11$ open-source systems and report that microservices frequently co-evolve, with certain services repeatedly changing together due to cross-cutting features and shared API concerns. Apolinário and de França~\cite{apolinario_method_2021} further quantify this phenomenon by proposing a method for monitoring the evolution of coupling in microservices using dependency-based metrics such as Service Dependency Distribution (SDD) and Service Interaction Density (SID). Their longitudinal analysis study of $17$ Spinnaker releases reveals a $28.6\%$ increase in dependencies and shows increasing coupling and dependency imbalance over time, indicative of hub-like~\cite{hub_like_azadi_2019} architectural anti-patterns. Similarly, Decan et al.~\cite{decan2019empirical} empirically analyse dependency networks across seven software package ecosystems and show that these networks evolve toward increasingly unequal, hub-dominated structures over time, with a small number of packages accumulating a disproportionate share of reverse dependencies. In a related study, Fritz et al.~\cite{georg2024strategic} model software dependency formation and show that creating a dependency on a particular package has a positive external effect on other developers. This effect increases the likelihood that additional dependencies will form on that package. Consequently, highly interdependent packages are likely to become even more interconnected.

These studies collectively establish that dependencies and their structural roles are both architecturally significant and dynamic. While they identify central services, hub-like structures, and temporal changes in dependency importance, they do not provide a theoretical understanding of how dependency concentration emerges during system evolution. In this work, we take a first step toward addressing this gap by proposing an initial theory of the dependency black hole, which conceptualises the evolution of dependency concentration and serves as a foundation for future empirical refinement and extension.

\section{Conclusions}\label{sec:conclusion}

This paper introduced the \emph{dependency black hole} as a theory-building lens for explaining dependency concentration in large-scale microservice architectures. Grounded in an industrial case study of $267$ microservices and iteratively validated with practitioners, the theory characterises dependency networks in terms of a singularity of dependency magnets, an event horizon of services increasingly entangled with the core, and an accretion disk of lightly connected services.

As future work, we plan to study similar cases with an explicit time component, i.e., observe microservice dependency networks longitudinally across large-scale software development organisations to validate the generality of the theory. A key next step is to measure the \emph{dynamics} of nodes over time---whether, how, and when services approach the event horizon and transition into the singularity---and to operationalise early-warning indicators that can support proactive architectural governance.

\bibliographystyle{IEEEtran}
\balance
\bibliography{references}
\end{document}